\documentclass[twocolumn,twoside]{article}

\usepackage{graphicx}
\usepackage{hyperref}

\renewcommand{\thefootnote}{\fnsymbol{footnote}}

\begin{document}

\thispagestyle{plain}
\markboth{\rm \uppercase{On creation of scalar particles ... \hspace{111pt} }}
{\hspace*{202pt} \rm \uppercase{Pavlov}}

\twocolumn[
\begin{center}
{\LARGE \bf On creation of scalar particles with

\vspace{10pt}
Gauss-Bonnet type coupling to curvature

\vspace{10pt}
in Friedmann cosmological models${}^{\tiny 1}$}
\vspace{12pt}

{\Large {Yu. V. Pavlov}${}^*$} \vspace{12pt}

{\it Institute of Problems in Mechanical Engineering, Russian Acad. Sci.,\\
Bol'shoy pr. 61, St. Petersburg 199178, Russia;}
\vspace{4pt}

{\it A. Friedmann Laboratory for Theoretical Physics, St.\,Petersburg, Russia;}
\vspace{4pt}

{\it Copernicus Center for Interdisciplinary Studies, Krak\'{o}w, Poland}
\end{center}
\vspace{5pt}
 {\bf Abstract.}
    Calculations are presented for creation of massive and massless scalar
particles coupled to Gauss-Bonnet type curvature in Friedmann cosmological
models.
    It has been shown that, for fields of mass~$m$, the effect of the coupling
parameter~$\zeta$ with the Gauss-Bonnet invariant is insignificant
if $\zeta m^2 \ll 1 $.
    In all cases under consideration, the created particle number is
compatible by order of magnitude with the number of causally disconnected
space-time regions by the Compton time, corresponding  $1/m$ or $\sqrt{\zeta}$.

\vspace{11pt}
{PACS number:} 04.62.+v
\vspace{17pt}
]

%%%%******************************************************************
{\centering  \section{Introduction}}

\footnotetext[1]{yuri.pavlov@mail.ru}
\footnotesep=9pt

\renewcommand{\thefootnote}{\arabic{footnote}}
\footnotetext[1]{Contribution to the issue of ``Gravitation and Cosmology''
dedicated to Prof. Andrei Grib's 75th birthday.}

    Quantum effects in the gravitational field have been actively studied
since the 70s of the last century.
    Quantum field effects, in particular, particle creation can have important
applications in the early Universe cosmology and astrophysics~\cite{GMM,BD}.
    A basic contribution to this branch of theoretical physics has been
made by A.A.~Grib and his students.
    In~\cite{GribNuclPhys69}, A.A.~Grib and S.G.~Mamayev have suggested
the Hamiltonian diagonalization method to describe particle creation by
the gravitational field.
    For the first time, they obtained finite expressions for the particle
number density created in homogeneous and isotropic models of the Universe.
    A great significance of the approach suggested in~\cite{GribNuclPhys69}
is connected with the definitions of the notions of a vacuum and an elementary
particle in curved space-time, where, in general, there is no symmetry group
like the Poincar\'{e} group in Minkowski space.
    If one assumes that a particle is associated with a quantum of energy,
then observation of a particle at a certain time instant means, according to
the principles of quantum theory, finding an eigenstate of the Hamiltonian.
    That is what is taken into account in the Hamiltonian diagonalization
method~\cite{GMM}.

    Numerical estimates have shown that the effect of particle creation with
masses of the order of the proton mass is negligibly small in cosmology.
    However, if the particle masses are large, e.g., those of the order of the
Grand Unification scale, then the number of particles created in the Friedmann
Universe will be of the order of the Eddington-Dirac number
$(\approx 10^{80})$  \cite{Grib95,GribDorofeev94}.
    This can be used in order to explain the observed baryonic charge of
the Universe and the origin of ultra-high energy cosmic
rays~\cite{GribPavlov2002(IJMPD)}--\cite{GrPvAGN}.

    Up to a certain time, the particle creation theory considered only
the case of conformally coupled fields.
    For fields with more general couplings with the curvature it was shown,
e.g., in~\cite{BMR,BLMPv}, that the corresponding nonconformal additions
can be dominant in both the particle creation effect and in the vacuum
mean values of the stress-energy tensor components.
    For nonconformally coupled fields, the metric Hamiltonian diagonalization
method led to infinite values of the created particle number~\cite{Fulling79}.
    This problem has been solved in~\cite{Pv2001,PvIJA}.
    Calculations of nonconformal particle creation with minimal coupling to
the curvature in Friedmann universes are presented in~\cite{GribPavlov2008}.

    In the present paper we consider the case of nonconformal coupling with
the Gauss-Bonnet type curvature.
    This type of scalar field coupling to gravity is a generalization of
the usual nonconformal coupling that does not change the kinetic term and
preserves the most important property of a scalar field: its metric
stress-energy tensor in this case also does not contain higher than
second-order derivatives of the metric and the field itself.
    For a model metric, admitting an exact analytical solution,
we have previously shown~\cite{Pavlov2013} that the existence of a
coupling with the Gauss–Bonnet invariant can exert a dominant influence in
the particle creation effect by the gravitational field.
    In the present paper we study particle creation in Friedmann cosmological
models for particles coupled with the Gauss-Bonnet type curvature.

    We use the system of units in which $\hbar \!=\! c \!=\! 1$.
    The signs of the Riemann and Ricci tensors are chosen so that
$ R^{\, i}_{\ jkl} = \partial_l \, \Gamma^{\, i}_{\, jk} -
\partial_k \, \Gamma^{\, i}_{\, jl} +
\Gamma^{\, i}_{\, nl} \Gamma^{\, n}_{\, jk} -
\Gamma^{\, i}_{\, nk} \Gamma^{\, n}_{\, jl}  $,
$\ R_{ik} = R^{\, l}_{\ ilk}$,
where $\Gamma^{\, i}_{\, jk}$ are the Christoffel symbols.

\vspace{14pt}
%%%%*****************************************************************
{\centering \section{Scalar field with Gauss-Bonnet type coupling}
\label{sec2}
}

    Consider a complex scalar field $\varphi(x)$ of mass $m$
with the Lagrangian
    \begin{equation}
L(x)=\sqrt{|g|} \left[g^{ik}\partial_i\varphi^*\partial_k\varphi -
(m^2 \!+\! \xi R + \zeta R_{GB}^{\,2}) \varphi^* \varphi \right]
\label{Lag}
\end{equation}
     and the corresponding equation of motion
\begin{equation}
( \nabla^i \nabla_{\! i} + m^2 + \xi R + \zeta R_{GB}^{\,2})\, \varphi(x)=0 ,
\label{Eqm}
\end{equation}
    where ${\nabla}_{\! i}$ are covariant derivatives in
$N$-dimensional space-time with the metric $g_{ik}$,
$ g\!=\!{\rm det}(g_{ik})$, \, $R$ is the scalar curvature,
    \begin{equation}
R_{GB}^{\,2} \stackrel{\rm def}{=}
R_{lmpq} R^{\,lmpq} - 4 R_{lm} R^{\,lm} + R^2 ,
\label{RGB}
\end{equation}
    $\xi$ is a dimensionless constant while the constant $\zeta$ has
the dimension of the inverse mass squared: $\zeta = 1/M_\zeta^2$.
    The integrals of $R$ and $R_{GB}^{\,2}$ are proportional, in the dimensions
$N=2$ and 4, respectively, to the Euler characteristics of the manifolds,
which are, according to the Gauss-Bonnet theorem, topological invariants.
    Therefore the coupling to curvature of the form~(\ref{Lag}), is called,
at $\zeta \ne 0$, the Gauss–Bonnet type coupling.

    Eq.~(\ref{Eqm}) is conformally invariant if $m=0$, $\zeta = 0$ and
    \begin{equation}
\xi= \xi_c \stackrel{\rm def}{=} \frac{N-2}{4 (N-1)}
\label{conf}
\end{equation}
($\xi_c = 1/6 $ for $N=4$).
    Therefore the case $\xi = \xi_c$, $\zeta = 0$ is called conformal
coupling to the curvature.
    The case $\xi = 0 = \zeta$ corresponds to minimal coupling.

    For the Gauss–Bonnet type coupling, the metric stress-energy tensor (SET)
of a scalar field~(\ref{Lag}) does not contain derivatives of the metric
higher than second-order.
    Explicit expressions for the SET are given in~\cite{Pavlov2004}, where,
for the first time, quantization and renormalization of the vacuum mean values
of SET components were carried out for a scalar field with this type of
coupling in a homogeneous and isotropic space-time.
    Taking into account the possible scalar field coupling with
the Gauss-Bonnet invariant $R_{GB}^{\,2}$ can be significant for the early
Universe, in black holes studies, in calculations of boson stars parameters,
etc.

    We write the metric of a homogeneous and isotropic space-time in the form
    \begin{equation}
ds^2= d t^2- a^2(t)\, d l^2 = a^2(\eta)\,(d{\eta}^2 - d l^2) ,
\label{gik}
\end{equation}
    where $d l^2 $ is the metric of an $(N-1)$-dimensional space of constant
curvature $K=0, \pm 1 $.
    In the metric~(\ref{gik}), the expressions for the scalar curvature and
for the Gauss–Bonnet invariant are (see~\cite{Pavlov2004})
     \begin{equation}
R= a^{-2}(N-1) \left[ \, 2c'+(N-2)(c^2+K) \right],
\label{RRRR}
\end{equation}
       \begin{eqnarray}
R_{GB}^{\,2} &=& a^{-4} (N-1) (N-2) (N-3)\, (c^2 +K)
\nonumber \\[2pt]
&&\times  \left[ 4 c'+ (N\!-\!4) \, (c^2 +K) \right],
\label{RGBoi}
\end{eqnarray}
    where the prime denotes a derivative in the conformal time $\eta$, \,
$c=a'/a = \dot{a}(t)$.

    The Einstein equations
    \begin{equation}
R_{ik} - \frac{1}{2} g_{ik} R = - 8 \pi G\, T_{ik} ,
\label{GR70Ein}
\end{equation}
    for the metric~(\ref{gik}) have the form
    \begin{equation}
\frac{c^2 +K}{a^2} = \frac{16 \pi G\, \varepsilon }{(N-1)(N-2)} \,,
\label{GR70Ein1}
\end{equation}
    \begin{equation}
- \frac{1}{a^2} \biggl[\, c^{\, \prime} + \frac{N-3}{2} \left( c^2 +K \right)
\biggr] = \frac{8 \pi G p }{N-2} \,,
\label{GR70Ein2}
\end{equation}
    where $\varepsilon$ and $p$ are the energy density and pressure of
the background matter whose SET is
    \begin{equation}
T^i_k = {\rm diag}\, ( \varepsilon, -p, \ldots , -p\,).
\label{Tikdiag}
\end{equation}
    For spatially flat cosmologies $(K=0)$, in the case where the pressure
of the background matter is proportional to its energy density,
    \begin{equation}
p =w \varepsilon \,, \ \ \ \ \ w = {\rm const },
\label{GR70pw}
\end{equation}
    at $w > -(N\!-3)/(N\!-1)$ from~(\ref{GR70Ein1}), (\ref{GR70Ein2}) we obtain
    \begin{equation}
a = a_0\, t^{q} = a_1 \eta^{\beta},
\label{GR70ate}
\end{equation}
where $ t, \eta \in (0, \infty) $,
    \begin{equation}
q = \frac{2}{(N \!-1)(w+1)}, \ \ \  \beta = \frac{q}{1 - q},
\label{GR70beq}
\end{equation}
    \begin{equation}
t = \frac{a_1}{\beta+1} \, \eta^{\beta +1}\!, \ \ \ \ \
a_0 = a_1^{1/(\beta+1)} (\beta+1)^{\beta /(\beta+1)}\!.
\label{GR70teta}
\end{equation}
    In the range under consideration
    \begin{equation}
w \in \biggl( - \frac{N-3}{N-1}\,, \ +\infty \biggr)
\label{GR70wint}
\end{equation}
    the exponents in the scale factor change in the intervals
$q \in (0, 1)$,\ \ $\beta \in (0, +\infty)$.
    The values
    \begin{equation}
w =0, \ \ \ q=\frac{2}{N-1}, \ \ \ \beta=\frac{2}{N-3}
\label{dust}
\end{equation}
    correspond to dust background matter, and
    \begin{equation}
w =\frac{1}{N-1}, \ \ \ q=\frac{2}{N}, \ \ \ \beta=\frac{2}{N-2}
\label{radiat}
\end{equation}
    to a radiation-dominated epoch.

\vspace{14pt}
%%%%*****************************************************************
{\centering \section{Scalar particle creation in a homogeneous isotropic
space}
\label{sec3}
}

    To calculate the number of created particle pairs we use the Hamiltonian
diagonalization method~\cite{GMM}.
    The full set of solutions to Eq.~(\ref{Eqm}) in the metric~(\ref{gik})
can be found in the form
    \begin{equation}
\varphi(x) = a^{-(N-2)/2} (\eta)\, g_\lambda (\eta) \Phi_J ({\bf x}) ,
\label{fgf}
\end{equation}
    where
    \begin{equation}
g_\lambda''(\eta)+\Omega^2(\eta)\,g_\lambda(\eta)=0 ,
\label{gdd}
\end{equation}
       \begin{equation}
\Omega^2(\eta) = m^2 a^2 +\lambda^2 + (\xi - \xi_c) a^2 R +
\zeta a^2 R_{GB}^{\,2} ,
\label{Ome}
\end{equation}
     \begin{equation}
\Delta_{N-1}\,\Phi_J ({\bf x}) = - \biggl( \lambda^2 -
\Bigl( \frac{N-2}{2} \Bigr)^{\!2} K \biggr) \Phi_J  ({\bf x}),
\label{DFlF}
\end{equation}
$J$ being the set of indices (quantum numbers), numbering the eigenfunctions
of the Laplace-Beltrami operator $\Delta_{N-1}$ in ($N-1$)-dimensional space.

    According to the Hamiltonian diagonalization method,
the functions $g_\lambda(\eta)$ should satisfy the following initial
conditions~\cite{Pv2001,PvIJA}:
    \begin{equation}
g_\lambda'(\eta_0)=i\, \Omega(\eta_0)\, g_\lambda(\eta_0) \,, \ \
|g_\lambda(\eta_0)|= \Omega^{-1/2}(\eta_0).
\label{icg}
\end{equation}

    If a quantum scalar field is in a vacuum state at the time
instant $\eta_0$, then the density of particle pairs created by
the time $\eta $ can be calculated (for the metric with $K=0$)
by the formula~\cite{GMM}
    \begin{equation}
n(\eta) = \frac{B_N}{2 a^{N-1}} \int \limits_0^\infty \! S_\lambda(\eta)\,
\lambda^{N-2}\, d \lambda,
\label{nN}
\end{equation}
where $B_N=\left[2^{N\!-3} \pi^{(N \!-1)/2} \Gamma((N\!-1)/2) \right]^{-1}\!,$
\ $\Gamma(z)$ is the gamma function,
    \begin{equation}
S_\lambda(\eta) = \frac{\left| g'_\lambda (\eta ) -
i \Omega \, g_\lambda (\eta ) \right|^2}{4 \Omega } .
\label{Sgg}
\end{equation}
As shown in~\cite{Pv2001},\,\cite{PvIJA}, $ S_\lambda \sim \lambda^{-6} $
and the integral in~(\ref{nN}) converges at $N<7$.

    The formula for the number of particle pairs created in
the volume $a^{N-1}(t)$ by the time~$t$, in the case of the scale
factor~(\ref{GR70ate}), can be written in the form
    \begin{equation}
N(t) = \left(\frac{a(t_*)}{t_*} \right)^{\! N-1} b^{(0)}_q(t) ,
\label{NtC}
\end{equation}
    where $t_*$ is a certain fixed time instant.
    The superscript~$(0)$ indicates that it is scalar particle creation
that is considered.
    As follows from~(\ref{NtC}), $ b^{(0)}_{\,q}(t) / (1-q)^{N-1} $
is the proportionality factor between the created particle number and
the number of causally disconnected regions at the time $t_*$
after the Big Bang.

%%%%%%%%%%%%%%%%%%%%%%%%%%%%%%%%%%%%%%%%%%%%%%%%%%
    Some explanations are in order.
    Suppose that at the Big Bang time, from each spatial point, some signals
have been emitted, propagating with the speed of light.
    Let us find the distance~$D_H(t)$ (the cosmological horizon size) to which
these signals come from the emission point by the instant at which the age of
the Universe is equal to~$t$.
    The null geodesic equation~$ds=0$ in the metric~(\ref{gik})
reduces to~$ dt= \pm a(t)\,dl $, therefore,
    \begin{equation}
D_H(t) = a(t) \Delta l = a(t) \int \limits_0^t \frac{d \tau}{a(\tau)} \,.
\label{Horizon}
\end{equation}
    For the power-law scale factor~(\ref{GR70ate})
    \begin{equation}
D_H(t) = t / (1-q) \,.
\label{Horizq}
\end{equation}
    The number of causally disconnected regions in the volume~$ a^{N-1}(t) $
is equal to
    \begin{equation}
N_{c} = \left( \frac{a(t)}{D_H(t)} \right)^{N-1} \!\!\!= (1-q)^{N-1}\,
\left( \frac{a(t)}{t} \right)^{N-1} \!\!,
\label{HorizPrN}
\end{equation}
    which confirms the above interpretation of Eq.~(\ref{NtC}).
%%%%%%%%%%%%%%%%%%%%%%%%%%%%%%%%%%%%%%%%%%%%%%%%%%

    In the case of the usually considered coupling between the scalar field
and the curvature, $\xi R \varphi^* \varphi$, one uses as the time $t_*$
in Eq.~(\ref{NtC}) the Compton time $t_C=1/m$ for a massive scalar field.
    Let us note that for a conformally coupled scalar field the results of
calculations of the created particle number do not depend on the instant when
the initial conditions are imposed as well as on the instant when the created
particles are being observed if these times are much smaller and much larger
than the Compton time, respectively.
    Thus the coefficient $ b^{(0)}_{\,q}(t) $ for a conformally couples
field is time-independent at $mt \gg 1$.

    The relationship between the created particle number and the number of
causally disconnected regions in Friedmann cosmologies at Compton time
has been pointed out by Grib (see, e.g.,~\cite{Grib95}).
    The results of numerical calculations of the coefficient $ b^{(0)}_{\,q} $
in four-dimensional space-time for power-law scale factors are presented,
e.g., in~\cite{GribPavlov2008} (see also~\cite{KuzminTkachev99a}).

     For a nonconformal scalar field, the results can crucially depend on
the choice of the initial time instant because $\Omega^2(\eta)$
can grow without limits at fixed $\lambda$ due to the growth of $|R(t)|$
and $|R_{GB}^{\,2}(t)|$ as $t \to 0$
(see~(\ref{RRRR}), (\ref{RGBoi}), (\ref{Ome})).
    However, under the condition
    \begin{equation}
(\xi - \xi_c) R (t) + \zeta R_{GB}^{\,2}(t) < 0 , \ \ \ t \to 0,
\label{xiRzRGBm0}
\end{equation}
    the initial time $t_0$ in the particle creation problem is determined by
the requirement $\Omega^2(t) \ge 0$ for $t > t_0$ and any momenta $\lambda$.
    In the opposite case ($\Omega^2(t) < 0$ for some values of~$\lambda$),
in the presence of a scalar field self-interaction, the vacuum state would be
reconstructed similarly to the symmetry violation mechanism.
    Thus we choose the initial time $t_0$ from the equality
    \begin{equation}
m^2 + (\xi - \xi_c) R (t_0) + \zeta R_{GB}^{\,2}(t_0) = 0 .
\label{xiRzR00}
\end{equation}
    For a nonconformal scalar field such an approach to the particle creation
problem was suggested for the first time in~\cite{GribPavlov2008}, where
the calculation results are presented for a scalar field with minimal coupling
and power-law scale factors.

    In the present work, particle creation has been studied for
a 4D homogeneous and isotropic space-time with flat spatial sections ($K=0$).
    For the power-law scale factors~(\ref{GR70ate}), exact solutions to
Eq.~(\ref{gdd}) with Gauss–Bonnet type coupling are unknown.
    Therefore, numerical calculations have been conducted for such particle
creation, and their results are presented in the following plots.
%%%%%%%%%%%%%%%%%%%%%%%%%%%%%%%%%%%%%%%%%%%%%%%%%
    \begin{figure}[ht]
\centering
\includegraphics[width=77mm]{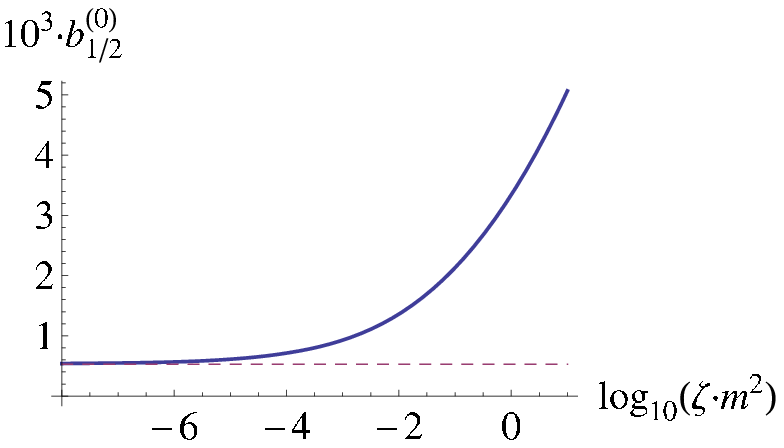}\\[-7pt]
\caption{Particle creation in a radiation-dominated Universe ($w=1/3$).}
\label{Grafg12}
\end{figure}

    Let us note that in the radiation-dominated case ($a=a_0 \sqrt{t}$),
presented in Fig.~\ref{Grafg12}, the scalar curvature $R=0$ and therefore
the value of the parameter~$\xi$ does not affect particle creation.
    The time $t_0$, according to~(\ref{xiRzR00}), is
    \begin{equation}
t_0 = \sqrt[{\textstyle 4}]{\frac{3 \zeta}{2 m^2}}.
\label{t0rd}
\end{equation}
    The dashed line in Fig.~\ref{Grafg12} corresponds to
the value $b^{(0)}_{1/2} = 5.3 \cdot 10^{-4}$ for a conformally coupled scalar
field in a radiation-dominated Universe~\cite{MMStarobinsky}.

\begin{figure}[ht]
\centering
\includegraphics[width=77mm]{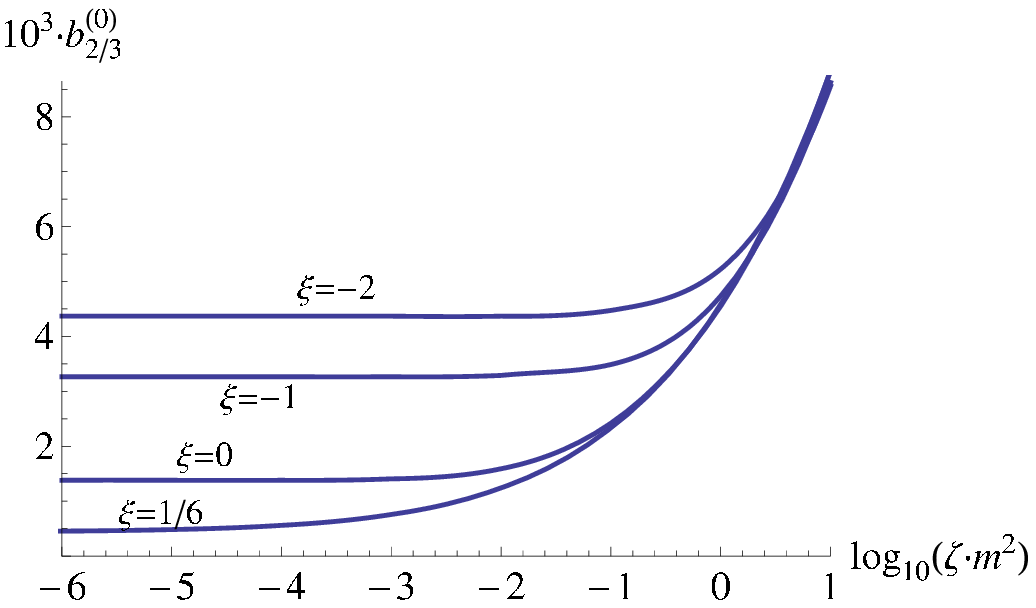}\\[-7pt]
\caption{Particle creation in a dust-dominated Universe ($p=0$).}
\label{Grafg23}
\end{figure}

    In the case of a dust-dominated Universe, $p=0$
(the scale factor $a=a_0 t^{2/3}$), the scalar curvature $R \ne 0$, and
the value of $\xi$ affects particle creation, as is evident
in Fig.~\ref{Grafg23}.
%%%%%%%%%%%%%%%%%%%%%%%%%%%%%%%%%%%%%%%%%%%%%%%%%

    In both Figs.~\ref{Grafg12} and~\ref{Grafg23}, the parameter $b^{(0)}_q$
was determined relative to the time $t_*=1/m$ (see Eq.~(\ref{NtC})).
    The value of $b^{(0)}_q$ was practically independent of the specific choice
of the final instant $t$ in the case $t \gg 1/m$.

    As is evident from Figs.~\ref{Grafg12} and~\ref{Grafg23}, the influence
of the coupling parameter $\zeta$ with the Gauss–Bonnet invariant on particle
creation is insignificant if $\zeta \ll 1/m^2$.
    The number of created particles in both cases, at $\zeta \le 1/m^2$,
is comparable by order of magnitude with the number of causally disconnected
regions in the corresponding Friedmann model by the Compton time ($t_C=1/m$)
for the scalar field mass $m$ from the beginning of the expansion.

\begin{figure}[ht]
\centering
\includegraphics[width=77mm]{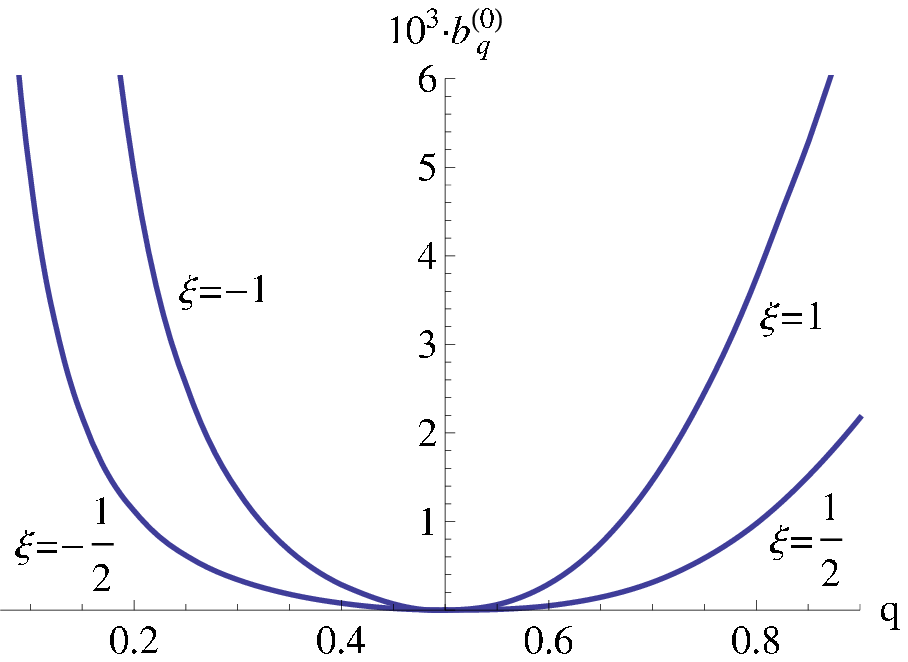}\\[-7pt]
\caption{Massless scalar particle creation for power-law scale factors.}
\label{Grafg4}
\end{figure}

    For a massless scalar field, the time $t_*$ in Eq.~(\ref{NtC}) was chosen
as  $t_* = \sqrt{\zeta}$, i.e., the Compton time for the mass scale of
the coupling parameter $\zeta$ between the scalar field and the Gauss-Bonnet
invariant~$R_{GB}^{\,2}$.
    In the massless case, the results of calculations of particle creation
for different values of the exponent $q$ of the scale factor~(\ref{GR70ate})
are presented in Fig.~\ref{Grafg4}.
%%%%%%%%%%%%%%%%%%%%%%%%%%%%%%%%%%%%%%%%%%%%%%%%%

    To provide the conditions~(\ref{xiRzRGBm0}), (\ref{xiRzR00}),
it was assumed
    \begin{equation}
\zeta >0 , \ \ \ \ (\xi - \xi_c ) (2 q -1) >0,
\label{moxize}
\end{equation}
    \begin{equation}
t_0 = 2 q \sqrt{\frac{\zeta (1-q)}{(\xi - \xi_c) (2q-1)}}.
\label{moxizet0}
\end{equation}
    The whole~$\zeta$ dependence of the created particle number is determined
in this case by the factor $( a(t_*)/t_* )^3$ in Eq.~(\ref{NtC})
and is proportional to $\zeta^{-3 (1-q)}$ (see Eq.~(\ref{HorizPrN})).
    The decrease in the particle number with growing $\zeta$ is explained in
this case by the growth of the time value $t_0$, at which the initial vacuum
state is postulated.
    As time grows, the gravitational field becomes weaker and creates
a smaller number of particles.

    In all cases considered (see Figs.~\ref{Grafg12}--\ref{Grafg4}),
the coefficient $b^{(0)}_q$ has the order $\approx 10^{-4}$ --- $10^{-2}$.
    Therefore, as in the case of conformally coupled particles in Friedmann
models, the number of particles created by the gravitational field of
the expanding Universe is comparable with the number of causally disconnected
regions by the Compton time~$t_C$ from the expansion beginning.
    For a massive field, at $\zeta m^2 \le 1$, this time is determined by
the field mass, $t_C=1/m$.
    For a massless field, this time corresponds to the mass scale of
the Gauss-Bonnet coupling parameter $t_C= \sqrt{\zeta}$.

\vspace{4pt}
{\centering \section*{Acknowledgment}}

This work were supported by a grant from the John Templeton Foundation.

\vspace{11pt}
%%%%%%%%%%%%%%%%%%%%%%%%%%%%%%%%%%%%%%%%%%%%%%%%%%%%%%%%%%%%%%%%%%%%%%


\begin{thebibliography}{99}
% \itemsep=0mm

\bibitem{GMM}
A.\,A.\,Grib, S.\,G.\,Mamayev, and V.\,M.\,Moste\-panenko,
{\it Vacuum Quantum Effects in Strong Fields}
(Energoatomizdat, Moscow, 1988, in Russian; English
translation: Friedmann Lab. Publ., St. Petersburg,
1994).

\bibitem{BD}
N.\,D. Birrell and P.\,C.\,W. Davies,
{\it Quantum Fields in Curved Space}
(Cambridge Univ. Press, Cambridge, 1982).

\bibitem{GribNuclPhys69}
A.\,A. Grib and S.\,G. Mamayev,
Yadernaya Fizika {\bf 10}, 1276 (1969)
[English transl.: Sov. J. Nucl. Phys. (USA) {\bf 10}, 722 (1970)].

\bibitem{Grib95}
A.\,A. Grib,
{\it Early Expanding Universe and Elementary Particles}
(Friedmann Lab. Publ., St.Petersburg, 1995).

\bibitem{GribDorofeev94}
A.\,A. Grib and V.\,Yu. Dorofeev,
Int. J. Mod. Phys.
\href{http://dx.doi.org/10.1142/S0218271894000848}
{D {\bf 3}, 731 (1994)}.

\bibitem{GribPavlov2002(IJMPD)}
A. A. Grib and Yu. V. Pavlov,
Int. J. Mod. Phys.
\href{http://dx.doi.org/10.1142/S0218271802001706}
{D {\bf 11}, 433} (2002).

\bibitem{GribPavlov2002(IJMPA)}
A. A. Grib and Yu. V. Pavlov,
Int. J. Mod. Phys.
\href{http://dx.doi.org/10.1142/S0217751X02013514}
{A {\bf 17}, 4435} (2002).

\bibitem{GrPvAGN}
A. A. Grib and  Yu. V. Pavlov,
Mod. Phys. Lett.
\href{http://dx.doi.org/10.1142/S0217732308027072}
{A {\bf 23}, 1151} (2008).

\bibitem{BMR}
V.\,B. Bezerra, V.\,M. Mostepanenko, and C. Ro\-me\-ro,
\href{http://dx.doi.org/10.1142/S0217732397000145}
{Mod. Phys. Lett.~A {\bf 12}, 145}(1997).

\bibitem{BLMPv}
M. Bordag, J. Lindig, V.\,M. Mostepanenko, and Yu.\,V. Pavlov,
\href{http://dx.doi.org/10.1142/S0218271897000261}
{Int. J. Mod. Phys.~D {\bf 6}, 449} (1997).

\bibitem{Fulling79}
S.\,A. Fulling,
\href{http://dx.doi.org/10.1007/BF00756661}
{Gen. Relativ. Gravit. {\bf 10}, 807} (1979).

\bibitem{Pv2001}
Yu. V. Pavlov,
\href{http://dx.doi.org/10.4213/tmf418}
{Teor. Mat. Fiz. {\bf 126}, 115} (2001)
[English transl.:
\href{http://dx.doi.org/10.1023/A:1005206315688}
{Theor. Math. Phys. {\bf 126}, 92} (2001)].

\bibitem{PvIJA}
Yu. V. Pavlov,
\href{http://dx.doi.org/10.1142/S0217751X02010479}
{Int. J. Mod. Phys.~A {\bf 17}, 1041} (2002).

\bibitem{GribPavlov2008}
A. A. Grib and Yu. V. Pavlov, Grav. Cosmol.
\href{http://dx.doi.org/10.1134/S0202289308010015}
{{\bf 14}, 1 (2008)}.

\bibitem{Pavlov2013}
Yu. V. Pavlov,
\href{http://dx.doi.org/10.4213/tmf8393}
{Teor. Mat. Fiz. {\bf 174}, 504} (2013)
[English transl.:
\href{http://dx.doi.org/10.1007/s11232-013-0036-y}
{Theor. Math. Phys. {\bf 174}, 438} (2013)].

\bibitem{Pavlov2004}
Yu. V. Pavlov,
\href{http://dx.doi.org/10.4213/tmf96}
{Teor. Mat. Fiz. {\bf 140} 241} (2004)
[English transl.:
\href{http://dx.doi.org/10.1023/B:TAMP.0000036540.50125.0c}
{Theor. Math. Phys. {\bf 140}, 1095} (2004)].

\bibitem{KuzminTkachev99a}
V. Kuzmin and I. Tkachev, Phys. Rev.
\href{http://dx.doi.org/10.1103/PhysRevD.59.123006}
{D {\bf 59}, 123006} (1999).

\bibitem{MMStarobinsky}
S.\,G.~Mamaev, V.\,M.~Mostepanenko, and A.\,A.~Starobinskii,
\href{http://jetp.ac.ru/cgi-bin/index/r/70/5/p1577?a=list}
{ZhETF {\bf 70},  1577} (1976)
\ [English transl.:
\href{http://jetp.ac.ru/cgi-bin/index/e/43/5/p823?a=list}
{Sov. Phys.--JETP {\bf 43}, 823} (1976)].

\end{thebibliography}
\end{document}